\documentclass{ws-procs9x6}

\newcommand{\intx}{\ensuremath{\int d^4 x}}
\newcommand{\sgn}{\ensuremath{\mathrm{sgn}}}
\newcommand{\intq}{\ensuremath{\int\frac{d^4 q}{(2\pi)^4}}}
\newcommand{\ppara}{\ensuremath{p_\parallel}}
\newcommand{\qpara}{\ensuremath{q_\parallel}}
\newcommand{\qperp}{\ensuremath{q_\perp}}
\newcommand{\gammapara}{\ensuremath{\gamma_\parallel}}
\newcommand{\nn}{\nonumber}
\begin{document}

\title{PROGRESS ON CHIRAL SYMMETRY BREAKING IN A STRONG MAGNETIC FIELD}

\author{SHANG-YUNG WANG}

\address{Department of Physics, Tamkang University,\\ Tamsui,
Taipei 25137, Taiwan\\
E-mail: sywang@taos.phys.tku.edu.tw}

\begin{abstract}
The problem of chiral symmetry breaking in QED in a strong magnetic
field is briefly reviewed. Recent progress on issues regarding the
gauge fixing independence of the dynamically generated fermion mass
is discussed.
\end{abstract}

%\keywords{Gauge independence; Chiral symmetry; Magnetic field; Schwinger--Dyson equations}

\vspace{0.25in}

\bodymatter

Gauge theories play an important role in our understanding of a wide
variety of phenomena in many areas of physics, ranging from the
descriptions of fundamental interactions in elementary particle
physics to the study of high temperature superconductivity in
condensed matter physics. While the usual perturbative approach
based on the loop expansion is sufficient in most circumstances,
there are many interesting phenomena that can be understood only
through a nonperturbative analysis. Examples include strong coupling
gauge theories (such as QCD and hadron physics) as well as gauge
fields under the influence of extreme conditions (such as high
temperature and/or high density, strongly out-of-equilibrium
instabilities, and strong external fields).

Several methods have been developed to study nonperturbative
phenomena in gauge theories. Lattice approach based on discretized
descriptions is well suited for studying strong coupling gauge
theories, especially the properties in vacuum or at finite
temperature. New approach based on the Anti-de Sitter/conformal
field theory (AdS/CFT) correspondence offers the possibility to
study strong coupling gauge theories by performing perturbative
calculations in their gravity duals. Yet, field theoretical
continuum approaches such as the Schwinger--Dyson (SD) equations and
the effective action provide a natural framework for studies of
nonperturbative phenomena in gauge theories. In particular,
continuum descriptions are capable of describing dynamical real-time
quantities in gauge fields under the influence of extreme
conditions. A consistent formulation with gauge fields is important
to the development of approximation techniques that will complement
investigations utilizing lattice gauge theory and the AdS/CFT
correspondence.

In this contribution we focus on the SD equations approach and
discuss the important issues regarding the consistency and gauge
independence of the truncation schemes therein. The SD equations are
an infinite set of coupled integral equations among the Green's
functions in a field theory, and form equations of motion of the
corresponding theory. Nevertheless, practical calculations
necessitate the use of approximated, or truncated, versions of the
exact SD equations. A truncation of the SD equations corresponds
diagrammatically to a selected resummation of an infinite subset of
diagrams arising from every order in the loop expansion. Hence, in
gauge theories the gauge independence of physical observables cannot
be guaranteed unless consistent schemes are employed in truncating
the SD equations.

We take as an example the simplest bare vertex approximation (BVA)
to the SD equations in the simplest gauge theory, i.e., QED with
massless Dirac fermions, in the presence of a background gauge
field. Specifically, we consider the problem of chiral symmetry
breaking in QED in a strong, external magnetic field. This problem
was originally motivated by a proposal~\cite{Caldi:1987pm} to
explain the correlated $e^+e^-$ peaks observed in heavy ion
collision experiments, and has subsequently received a lot of
attention over the past decade largely because of its applications
in astrophysics, condensed matter physics and cosmology.

Since the dynamics of fermion pairing in a strong magnetic field is
dominated by the lowest Landau level (LLL)~\cite{Gusynin:1994re}, it
is a common practice to consider the propagation of, as well as
radiative corrections originating only from, fermions occupying the
LLL. This is referred to in the literature as the lowest Landau
level approximation
(LLLA)~\cite{Gusynin:1994re,Gusynin:1995gt,Gusynin:1998zq}. In
Ref.~\refcite{Gusynin:1995gt} chiral symmetry breaking in QED in a
strong magnetic field was first studied in the so-called (quenched)
rainbow approximation, in which the bare vertex and the free photon
propagator were used in truncating the SD equations. These studies
provided a preliminary affirmation of the phenomenon of chiral
symmetry breaking in QED in a strong magnetic field.

More recently, this phenomenon has been studied in the so-called
improved rainbow
approximation~\cite{Gusynin:1998zq,Kuznetsov:2002zq}, in which the
bare vertex and the full photon propagator were used. The authors of
Ref.~\refcite{Gusynin:1998zq} claimed that (i) in covariant gauges
there are one-loop vertex corrections arising from the longitudinal
components of the full photon propagator that are not suppressed by
powers of the gauge coupling constant and hence need to be accounted
for; (ii) there exists a noncovariant and nonlocal gauge in which,
and only in which, the BVA is a reliable truncation of the SD
equations that consistently resums these one-loop vertex
corrections; (iii) the phenomenon of chiral symmetry breaking is
universal in that it takes place for any number of the fermion
flavors. In Ref.~\refcite{Kuznetsov:2002zq} the authors included
contributions to the vacuum polarization from higher Landau levels
in an unspecified gauge, and asserted that (i) in QED with $N_f$
fermion flavors a critical number $N_{cr}$ exists for any value of
the gauge coupling constant, such that chiral symmetry remains
unbroken for $N_f>N_{cr}$; (ii) the dynamical fermion mass is
generated with a double splitting for $N_f<N_{cr}$. The results of
Refs.~\refcite{Gusynin:1998zq,Kuznetsov:2002zq} are clearly in
contradiction, leading to a controversy~\cite{Gusynin:2002yi} over
the correct calculation of the dynamical fermion mass generated
through chiral symmetry breaking in a strong magnetic field.

The controversy was later resolved in Ref.~\refcite{Leung:2005yq} by
establishing the gauge independence of the dynamical fermion mass
calculated in the SD equations approach. In particular, it was shown
that (i) the BVA is a consistent truncation of the SD equations in
the LLLA; (ii) within this consistent truncation scheme the physical
dynamical fermion mass, obtained as the solution of the truncated
fermion SD equations evaluated on the mass shell, is manifestly
gauge independent.

We take the constant external magnetic field of strength $H$ in the
$x_3$-direction. The corresponding vector potential is given by
$A_\mu=(0,0,Hx_1,0)$, where $\mu=0,1,2,3$. In our convention, the
metric has the signature $g_{\mu\nu}=\mathrm{diag}(-1,1,1,1)$. In
the LLLA, the SD equations for the full fermion propagator $G(x,y)$
are given by
\begin{eqnarray}
G^{-1}(x,y)&=&S^{-1}(x,y)+\Sigma(x,y),\label{SDfermion}\\
\Sigma(x,y)&=&ie^2\intx' d^4y' \,\gamma^\mu\,G(x,x')\,
\Gamma^\nu(x',y,y')\,\mathcal{D}_{\mu\nu}(x,y'),\label{SDSigma}
\end{eqnarray}
where $S(x,y)$ is the bare propagator for the LLL fermion in the
external field $A_\mu$, $\Sigma(x,y)$ is the LLL fermion self-energy, and
$\Gamma^\nu(x,y,z)$ is the full LLL fermion-photon vertex. The full
photon propagator $\mathcal{D}_{\mu\nu}(x,y)$ satisfies the SD
equations
\begin{eqnarray}
\mathcal{D}^{-1}_{\mu\nu}(x,y)&=&D^{-1}_{\mu\nu}(x,y)+\Pi_{\mu\nu}(x,y),\\
\Pi_{\mu\nu}(x,y)&=&-ie^2\,\mathrm{tr}\intx' d^4y'\,\gamma_\mu\,G(x,x')\,
\Gamma_\nu(x',y',y)\,G(y',x),\label{SDphoton}
\end{eqnarray}
where $D_{\mu\nu}(x,y)$ is the free photon propagator and
$\Pi_{\mu\nu}(x,y)$ is the vacuum polarization. In the BVA to the SD
equation one replaces the full vertex by the bare one, viz,
$\Gamma^\mu(x,y,z)=\gamma^\mu\,\delta^{(4)}(x-z)\,\delta^{(4)}(y-z)$.

A consistent truncation of the SD equations is that which respects
the Ward--Takahashi (WT) identity satisfied by the truncated vertex
and inverse fermion propagator. The WT identity in the BVA within
the LLLA was first studied in Ref.~\refcite{Ferrer:1998vw}. It was
shown that in order to satisfy the WT identity in the BVA within the
LLLA, the LLL fermion self-energy in momentum space has to be a
momentum independent constant. As per the WT identity in the BVA,
the LLL fermion self-energy takes the form $\Sigma(\ppara)=m_\xi$,
where $\ppara$ is the momentum of the LLL fermion and $m_\xi$ is a
momentum independent \emph{but gauge dependent} constant, with $\xi$
being the gauge fixing parameter. Here and henceforth, the subscript
$\parallel$ ($\perp$) refers to the longitudinal: $\mu=0,3$
(transverse: $\mu=1,2$) components. It is noted that $m_\xi$ depends
implicitly on $\xi$ through the full photon propagator
$\mathcal{D}_{\mu\nu}$ in \eqref{SDSigma}. We emphasize that because
of its $\xi$-dependence, $m_\xi$ should \emph{not} be taken for
granted to be the dynamical fermion mass, which is a gauge
independent physical observable. This is one of the subtle points
that has been overlooked in the literature.

We now show that the BVA within the LLLA is a consistent truncation
of the SD equations \eqref{SDfermion}--\eqref{SDphoton}, in which
$m_\xi$ is $\xi$-independent and hence can be identified
unambiguously as the physical dynamical fermion mass if, and only
if, the truncated SD equation for the fermion self-energy is
evaluated on the fermion mass shell. First recall that, as proved in
Ref.~\refcite{Kobes:1990dc}, in gauge theories the singularity
structures (i.e., the positions of poles and branch singularities)
of gauge boson and fermion propagators are gauge independent when
all contributions of a given order of a systematic expansion scheme
are accounted for. Consequently, the physical dynamical fermion mass
has to be determined by the pole of the full fermion propagator
obtained in a consistent truncation scheme of the SD equations.
Assume for the moment that the BVA is a consistent truncation in the
LLLA, such that the position of the pole of the LLL fermion
propagator is gauge independent. In accordance with the WT identity
in the BVA, we have
\begin{equation}
\Sigma(\ppara)=m,\label{Sigma}
\end{equation}
where the constant $m$ is the gauge independent, physical dynamical
fermion mass, yet to be determined by solving the truncated SD
equations self-consistently. What remains to be verified is the
following statements: (i) the truncated vacuum polarization is
transverse; (ii) the truncated fermion self-energy is gauge
independent when evaluated on the mass shell, $\ppara^2=-m^2$. We
highlight that the fermion mass shell condition is one of the
important points that has gone unnoticed in the literature, where
the truncated fermion self-energy used to be evaluated off the mass
shell at
$\ppara^2=0$~\cite{Gusynin:1995gt,Gusynin:1998zq,Kuznetsov:2002zq}.

The vacuum polarization $\Pi_{\mu\nu}(q)$ in the BVA is found to be
given by
\begin{eqnarray}
\Pi^{\mu\nu}(q)&=&-\frac{ie^2}{2\pi}\,N_f\,|eH|\,
\exp\left(-\frac{q_\perp^2}{2|eH|}\right)
\mathrm{tr}\int\frac{d^2p_\parallel}{(2\pi)^2}
\gamma^\mu_\parallel\,\frac{1}{\gammapara\cdot\ppara+m}\,
\gamma^\nu_\parallel\nn\\
&&\times\frac{1}{\gammapara\cdot(p-q)_\parallel+m}\,\Delta[\sgn(eH)],
\label{Pi}
\end{eqnarray}
where $\Delta[\sgn(eH)]=[1+i\gamma^1\gamma^2\,\sgn(eH)]/2$ is the
projection operator on the fermion states with the spin parallel to
the external magnetic field. The presence of $\Delta[\sgn(eH)]$ in
\eqref{Pi} is a consequence of the LLLA, which implies an effective
dimensional reduction from $(3+1)$ to $(1+1)$ in the fermion
sector~\cite{Gusynin:1995gt}.

The WT identity in the BVA guarantees that the vacuum polarization
$\Pi_{\mu\nu}(q)$ is transverse, viz, $q^\mu \Pi_{\mu\nu}(q)=0$.
An explicit calculation yields
$\Pi^{\mu\nu}(q)=\Pi(\qpara^2,\qperp^2)
(g_\parallel^{\mu\nu}-q^\mu_\parallel q^\nu_\parallel/\qpara^2)$,
which in turn implies that the full photon propagator takes the
following form in covariant gauges ($\xi=1$ is the Feynman gauge):
\begin{equation}
\mathcal{D}^{\mu\nu}(q)=\frac{1}{q^2+\Pi(\qpara^2,\qperp^2)}
\left(g_\parallel^{\mu\nu}-\frac{q^\mu_\parallel
q^\nu_\parallel}{\qpara^2}\right)+\frac{g_\perp^{\mu\nu}}{q^2}
+\frac{q^\mu_\parallel q^\nu_\parallel}{q^2 \qpara^2}
+(\xi-1)\frac{1}{q^2}\frac{q^\mu q^\nu}{q^2}. \label{D}
\end{equation}
In the above expressions, the polarization function
$\Pi(\qpara^2,\qperp^2)$ is given by
\begin{equation}
\Pi(\qpara^2,\qperp^2)=\frac{2\alpha}{\pi}\,N_f\,
|eH|\,\exp\left(-\frac{q_\perp^2}{2|eH|}\right)
F\left(\frac{\qpara^2}{4m^2}\right),
\end{equation}
where $\alpha=e^2/4\pi$. The function $F(u)$ has the following
asymptotic behavior: $F(u)\simeq 0$ for $|u|\ll 1$ and $F(u)\simeq
1$ for $|u|\gg 1$. This implies that photons of momenta
$m^2\ll|\qpara^2|\ll|eH|$ and $\qperp^2 \ll |eH|$ are screened with
a characteristic length $L=(2\alpha N_f|eH|/\pi)^{-1/2}$. This
screening effect renders the rainbow
approximation~\cite{Gusynin:1995gt} completely unreliable in this
problem.

The fermion self-energy in the BVA, when evaluated on the fermion
mass shell, $\ppara^2=-m^2$, is given by
\begin{eqnarray}
m\,\Delta[\sgn(eH)]&=&ie^2\intq\,\exp\left(-\frac{q_\perp^2}{2|eH|}\right)
\gammapara^\mu\,\frac{1}{\gammapara\cdot(p-q)_\parallel+m}
\,\gammapara^\nu\,\nn\\
&&\times\mathcal{D}_{\mu\nu}(q)\,\Delta[\sgn(eH)]\bigg|_{\ppara^2=-m^2},
\label{m}
\end{eqnarray}
where $\mathcal{D}_{\mu\nu}(q)$ is given by \eqref{D}. The WT
identity in the BVA guarantees that this \emph{would-be} gauge
dependent contribution to the fermion self-energy (denoted
symbolically as $\Sigma_\xi$) is proportional to $(\gammapara \cdot
\ppara + m)$. We find through an explicit calculation that
\begin{align}
\Sigma_\xi&=\alpha\,(\xi-1)\,(\gammapara\cdot\ppara+m)\int_0^1
dx\,(1-x) \int\frac{d^2\qperp}{(2\pi)^2}\,
\exp\biggl(-\frac{q_\perp^2}{2|eH|}\biggr)\nn\\
&\quad\times\frac{(1-x)\qperp^2-xm(\gammapara\cdot\ppara-m)}{[(1-x)
\qperp^2+x(1-x)\ppara^2+xm^2]^2}\,\Delta[\sgn(eH)]. \label{mxi}
\end{align}
Hence, $\Sigma_\xi$ vanishes identically on the fermion mass shell
$\ppara^2=-m^2$ or, equivalently, $\gammapara\cdot\ppara+m=0$. This,
together with the transversality of the vacuum polarization,
completes our proof that the BVA is a consistent truncation of the
SD equations. Consequently, the dynamical fermion mass, obtained as
the solution of the truncated fermion SD equations evaluated on the
fermion mass shell, is gauge independent.

It can be verified in a similar manner that contributions to the
fermion self-energy from the longitudinal components in
$\mathcal{D}^{\mu\nu}(q)$ that are proportional to $q^\mu_\parallel
q^\nu_\parallel/\qpara^2$ also vanish when evaluated on the fermion
mass shell. Therefore, only the first term in
$\mathcal{D}^{\mu\nu}(q)$ proportional to $g_\parallel^{\mu\nu}$
contributes to the on-shell fermion self-energy. As a result, the
matrix structures on both sides of \eqref{m} are consistent. Using
the mass shell condition $\ppara^\mu=(m,0)$, corresponding to a LLL
fermion at rest, we find \eqref{m} in Euclidean space to be given by
\begin{equation}
m=\frac{\alpha}{2\pi^2}\int
d^2\qpara\frac{m}{q_3^2+(q_4-m)^2+m^2}\int_0^\infty d\qperp^2
\frac{\exp(-q_\perp^2/2|eH|)}{\qpara^2+\qperp^2+\Pi(\qpara^2,\qperp^2)},
\label{gap}
\end{equation}
where $\qpara^2 = q_3^2 + q_4^2$. Numerical analysis shows that the
solution of \eqref{gap} can be fit by the following analytic
expression:
\begin{equation}
m=a\,\sqrt{2|eH|}\;\beta(\alpha)\,
\exp\left[-\frac{\pi}{\alpha\log(b/N_f\,\alpha)}\right],\label{mass}
\end{equation}
where $a$ is a constant of order one, $b\simeq 2.3$, and
$\beta(\alpha)\simeq N_f\,\alpha$. From \eqref{mass} it follows that
in a strong magnetic field chiral symmetry is broken regardless
of the number of the fermion flavors.

The results of Refs.~\refcite{Gusynin:1998zq,Kuznetsov:2002zq} can
be attributed to gauge dependent artifacts. Had the authors of
Ref.~\refcite{Gusynin:1998zq} calculated properly the on-shell,
physical dynamical fermion mass, they would not have found the
``large vertex corrections'' they obtained, and therefore their
claim that the BVA is a good approximation only in the special
noncovariant and nonlocal gauge they invoke is not valid. In fact,
that special gauge was invoked by hand such that the gauge dependent
contribution cancels contributions from terms proportional to
$\qpara^\mu \qpara^\nu/\qpara^2$ in $\mathcal{D}^{\mu\nu}(q)$. Our
gauge independent analysis in the BVA reveals clearly that such a
gauge fixing not only is ad hoc and unnecessary, but also leaves the
issue of gauge independence unaddressed.

The truncation used in Ref.~\refcite{Kuznetsov:2002zq} is \emph{not}
a consistent truncation of the SD equations because the WT identity
in the BVA can be satisfied only within the
LLLA~\cite{Leung:2005yq}. Their result suggests that in the
inconsistent truncation as well as in the unspecified gauge, the
gauge dependent unphysical contributions from higher Landau levels
become dominant over the gauge independent physical contribution
from the LLL, thus leading to the authors' incorrect conclusions.
Therefore we emphasize that the LLL dominance in a strong magnetic
field should be understood in the context of a consistent
truncation. Namely, contributions to the dynamical fermion mass from
higher Landau levels that are obtained in a (yet to be determined)
consistent truncation of the SD equations are subleading when
compared to that from the LLL obtained in the consistent BVA
truncation. Research along this line will be the subject of further
investigations.

I would like to thank the organizers of the conference for their
invitation and hospitality. I am grateful to C.\ N.\ Leung for a
careful reading of the manuscript. This work was done in
collaboration with C.\ N.\ Leung, and was supported in part by the
National Science Council of Taiwan.

\end{document}